\documentclass[%
 reprint,
 amsmath,amssymb,
 aps,
]{revtex4-2}

\usepackage{amsmath}
\usepackage{quantikz}
\usepackage{hyperref} 
\usepackage{lineno}
\usepackage{graphicx}
\usepackage{slashed}
\usepackage{dcolumn}
\usepackage{bm}
\usepackage{booktabs} 
\usepackage{siunitx}  
\usepackage{amsmath}  
\usepackage{float}
\usepackage{titlesec}
\titlespacing*{\section}{10pt}{*2}{2.7ex}

\begin{document}


\preprint{APS/123-QED}

\title{Minimal length effect  on meson form factors in light front \texorpdfstring{AdS$_{5}$/QCD}{AdS5/QCD}}
\thanks{A footnote to the article title}%

\author{Fidele J. Twagirayezu}
 \altaffiliation{Department of Physics and Astronomy, University of California, Los Angeles.}
 \email{fjtwagirayezu@physics.ucla.edu}
\affiliation{Department of Physics and Astronomy University of California Los Angeles, Los Angeles, CA, 90095, USA\\
}%


\begin{abstract}
In this article, we investigate the impact of a minimal length scale, introduced via the Generalized Uncertainty Principle (GUP), on meson form factors within light-front holographic QCD (\(\mathrm{AdS}_5 / \mathrm{QCD}\)). By incorporating GUP through deformed operators in the QCD Lagrangian, we derive a GUP-corrected light-front wave function (LFWF) that includes contributions from all Fock states, weighted by probabilities \(a_n^2\). The resulting form factors account for transitions between Fock states via coefficients \(\mathcal{C}_{nm}\), revealing a net positive \(\beta\)-like correction driven by dominant positive coefficients (e.g., \(\mathcal{C}_{04}\), \(\mathcal{C}_{12}\)). This enhancement improves agreement with experimental pion form factor data. Our model is formulated to include all Fock states via the general summation over \( n \), but numerical evaluations truncate to \( n=0, 1, 2 \), sufficient for \( Q^2 \leq 4 \, \text{GeV}^2 \)~\eqref{tab:pion_form_factor_data}. Our findings suggest that minimal length effects amplify Fock state overlaps, offering insights into the interplay between quantum gravitational corrections and strong interaction dynamics. This generalized framework advances LFH QCD by capturing multi-parton contributions and paves the way for studying GUP effects in other hadronic systems.
\end{abstract}

\maketitle


\section{\label{sec:level1}Introduction} 
Quantum Chromodynamics (QCD) governs the strong interactions of quarks and gluons, yet its non-perturbative behavior at low energies challenges the understanding of hadron spectroscopy and dynamics. Light-front holographic QCD (LFH QCD), leveraging the AdS/CFT correspondence, maps strongly coupled QCD in four-dimensional Minkowski space to a weakly coupled gravitational theory in five-dimensional anti-de Sitter (AdS) space~\cite{brodsky_2015}. This framework simplifies confinement modeling and reproduces key hadronic features, such as linear Regge trajectories. At short distances, quantum gravitational effects may alter quantum mechanics, introducing a minimal length scale via the Generalized Uncertainty Principle (GUP)~\cite{Maggiore:1993kv, Kempf:1996nk, Capozziello:1999wx, Ali:2009zq, Ali_2011,ali2015minimallengthquantumgravity,Bosso:2023aht,Wagner:2023fmb,Twagirayezu:2020chx, Twagirayezu:2021ozk, Twagirayezu:2021kgh,Twagirayezu:2022uow,Twagirayezu:2022ucx}. By modifying the Heisenberg uncertainty principle, GUP regularizes ultraviolet divergences and influences QCD dynamics, particularly in LFH QCD, where the holographic coordinate corresponds to transverse momentum scales.
In this study, we explore the impact of GUP-induced minimal length on meson form factors within LFH QCD, generalizing the framework to include all Fock states weighted by probabilities \(a_n^2\). The  GUP is incorporated through deformed operators in the QCD Lagrangian, leading to a GUP-corrected light-front wave function (LFWF) that accounts for multi-parton contributions. The resulting form factors, modified by transition coefficients \(\mathcal{C}_{nm}\), exhibit a net positive \(\beta\)-like correction, enhancing agreement with experimental pion data. Our findings suggest that minimal length effects amplify Fock state overlaps, offering novel insights into the interplay between quantum gravitational effects and strong interactions. Light mesons, particularly pions, show heightened sensitivity to these corrections, highlighting their role as probes of fundamental physics.
The organization of this article is as follows: In Sec.~\eqref{sec:citeref}, we provide an overview for the derivation of the meson form factor in the standard light-front holographic QCD (LFH QCD). In Sec.~\eqref{sec:3yy}, we utilize the GUP-corrected light-front holographic wavefunction (LFHWF) derived in Ref.~\cite{Twagirayezu:2025tjc} to obtain the GUP-corrected meson form factors, including all Fock states. In Sec.~\eqref{sec:phen} 
we make plots of \(F(Q^2)\) and \(Q^2 F(Q^2)\), comparing theoretical predictions with and without GUP corrections to experimental data. In Sec.~\eqref{sec:concl5} we conclude our article with implications and future directions.
\section{\label{sec:citeref} Light Front Holographic QCD Theory}
The light-front holographic QCD theory connects Quantum Chromodynamics (QCD) to a holographic framework, specifically using the AdS/CFT correspondence, a strongly coupled QCD dynamics is mapped to a weakly coupled gravitational theory in a higher-dimensional anti-de Sitter (AdS) space. The process requires several steps to bridge the QCD Lagrangian to the light-front holographic QCD (LFH QCD) framework. Below, we provide a brief review for the derivation of relevant equation, focusing on the effective light-front Schrödinger-like equation for hadronic bound states, which is central to LFH QCD.
\subsection{QCD Lagrangian}
The QCD Lagrangian for \( N_f \) quark flavors and \( N_c = 3 \) colors is
\begin{equation}\label{eq:1x}
\begin{aligned}
\mathcal{L}_{\text{QCD}} = \sum_f \bar{\psi}_f (i\slashed{D}, - m_f) \psi_f - \frac{1}{4} G^a_{\mu\nu} G^{a\mu\nu},
\end{aligned}
\end{equation}
where \( \psi_f \) are quark fields for flavor \( f \), the quantity, \( D_\mu = \partial_\mu - i g A^a_\mu t^a \) is the covariant derivative, with gluon fields \( A^a_\mu \) and SU(3) generators \( t^a \), \( m_f \) is the quark mass, \( G^a_{\mu\nu} = \partial_\mu A^a_\nu - \partial_\nu A^a_\mu + g f^{abc} A^b_\mu A^c_\nu \) is the gluon field strength, and \( g \) is the strong coupling constant.
Mesons are quark-antiquark (\( q\bar{q} \)) bound states, and their form factors describe their response to external probes (e.g., electromagnetic currents). Computing form factors directly from this Lagrangian is intractable due to non-perturbative effects in the low-energy regime.
\subsection{Light-Front Quantization}
To make progress, we use light-front (LF) quantization, which simplifies the treatment of bound states. In LF coordinates, \( x^\pm = x^0 \pm x^3 \), \( \mathbf{x}_\perp = (x^1, x^2) \), the time coordinate is \( x^+ \), and the conjugate momentum is \( p^- = p^0 - p^3 \). The LF Hamiltonian is derived from the QCD Lagrangian by integrating out constrained fields (e.g., \( \psi_- \), parts of \( A_\mu \)) and quantizing on the light-front.
The QCD Hamiltonian in LF coordinates is complex due to gluon interactions, but for mesons, we focus on the \( q\bar{q} \) sector. The meson’s light-front wave function (LFWF) \( \psi(x, \mathbf{k}_\perp) \) describes the probability amplitude for finding a quark with longitudinal momentum fraction \( x = p_q^+ / P^+ \) and transverse momentum \( \mathbf{k}_\perp \), where \( P^+ \) is the meson’s total longitudinal momentum.
The LFWF satisfies a Schrödinger-like equation derived from the LF Hamiltonian
\begin{equation}
\begin{aligned}
H_{\text{LF}} | \psi \rangle = M^2 | \psi \rangle,
\end{aligned}
\end{equation}
where \( M \) is the meson mass, and \( H_{\text{LF}} \) includes kinetic terms and interactions. Solving this directly in QCD is still challenging due to non-perturbative dynamics.
\subsection{Transition to Holographic QCD}
To model non-perturbative QCD, LFH QCD uses the AdS/CFT correspondence, where a strongly coupled gauge theory (like QCD) is dual to a weakly coupled gravity theory in AdS\(_5\) space. The QCD Lagrangian does not directly yield the holographic dual, but we assume a correspondence where meson operators (e.g., \( \bar{q} \gamma^\mu q \)) map to fields in AdS.
In the soft-wall model, confinement is introduced via a dilaton field or potential in AdS. The AdS action for a scalar field \( \Phi(z) \) (dual to a meson) is
\begin{equation}
\begin{aligned}
S = \int d^4x \, dz \, \sqrt{g} \left[ g^{MN} \partial_M \Phi \partial_N \Phi + m_5^2 \Phi^2 + \kappa^2 z^2 \Phi^2 \right],
\end{aligned}
\end{equation}
where \( z \) is the holographic coordinate (inverse energy scale). \( g_{MN} \) is the AdS\(_5\) metric, \( m_5^2 \) is the mass related to the operator dimension (\( m_5^2 = \Delta(\Delta-4) \), with \( \Delta = 3 \) for mesons), and \( \kappa \) is the confinement scale from the soft-wall potential \( V(z) = \kappa^2 z^2 \).
Varying the action gives the equation of motion for the meson’s holographic wave function \( \phi(z) \)
\begin{equation}
\begin{aligned}
\left[ -\frac{d^2}{dz^2} + \frac{4L^2 - 1}{4z^2} + \kappa^2 z^2 \right] \phi(z) = M^2 \phi(z).
\end{aligned}
\end{equation}
\subsection{Light-Front Wave Function}
The holographic wave function \( \phi(z) \) is mapped to the LF wave function \( \psi(x, \mathbf{k}_\perp) \). In LFH QCD, the transverse separation \( z \) is related to the LF variables via \( z \sim \frac{\sqrt{x(1-x)}}{\kappa} \). The LFWF is obtained by Fourier transforming \( \phi(z) \)
\begin{equation}
\begin{aligned}
\psi(x, \mathbf{k}_\perp) =\frac{N}{\kappa}  \sqrt{x(1-x)} \exp\left( -\frac{\mathbf{k}_\perp^2}{2\kappa^2 x(1-x)} \right),
\end{aligned}
\end{equation}
where N is the normalization constant. We consider $N =2\sqrt{6\pi}$ so that 
the LFWF is normalized as
\begin{equation}
\begin{aligned}
\int_0^1 dx \int \frac{d^2 \mathbf{k}_\perp}{(2\pi)^2} |\psi(x, \mathbf{k}_\perp)|^2 = 1.
\end{aligned}
\end{equation}
The LFWF encodes the meson’s structure, derived indirectly from the QCD Lagrangian via the holographic correspondence.
\subsection{ Electromagnetic Form Factor}
The meson’s electromagnetic form factor \( F(Q^2) \) is defined by the matrix element of the electromagnetic current \( J^\mu = \sum_f e_f \bar{\psi}_f \gamma^\mu \psi_f \) as
\begin{equation}
\begin{aligned}
\langle P' | J^\mu | P \rangle = (P + P')^\mu F(Q^2),
\end{aligned}
\end{equation}
where \( Q^2 = -(P' - P)^2 \). In the LF framework, we use the Drell-Yan-West frame (\( q^+ = 0 \)), so \( Q^2 = \mathbf{q}_\perp^2 \), and compute \( J^+ \).
The form factor is the overlap of initial and final LFWFs
\begin{equation}\label{eq:8x}
\begin{aligned}
F(Q^2) = \sum_f e_f \int_0^1 dx \int \frac{d^2 \mathbf{k}_\perp}{16\pi^3} \psi^*(x, \mathbf{k}_\perp') \psi(x, \mathbf{k}_\perp),
\end{aligned}
\end{equation}
where \( \mathbf{k}_\perp' = \mathbf{k}_\perp + (1-x) \mathbf{q}_\perp \). Substituting the LFWF
\begin{equation}
\begin{aligned}
\psi(x, \mathbf{k}_\perp) = \frac{N}{\kappa} \sqrt{x(1-x)} \exp\left( -\frac{\mathbf{k}_\perp^2}{2\kappa^2 x(1-x)} \right),
\end{aligned}
\end{equation}
the integrand in Eq.~\eqref{eq:8x} becomes
\begin{equation}
\begin{aligned}
\psi^*(x, \mathbf{k}_\perp') \psi(x, \mathbf{k}_\perp) &= \frac{N^{2}}{\kappa^2} x(1-x)\\
&\times\exp\left( -\frac{\mathbf{k}_\perp^2 + (\mathbf{k}_\perp + (1-x) \mathbf{q}_\perp)^2}{2\kappa^2 x(1-x)} \right).
\end{aligned}
\end{equation}
The complete square in the exponent is
\begin{equation}
\begin{aligned}
\mathbf{k}_\perp^2 + (\mathbf{k}_\perp + (1-x) \mathbf{q}_\perp)^2 &= 2\left( \mathbf{k}_\perp + \frac{(1-x) \mathbf{q}_\perp}{2} \right)^2 \\
&+ \frac{(1-x)^2 \mathbf{q}_\perp^2}{2}.
\end{aligned}
\end{equation}
Thus, the form factor in Eq.~\eqref{eq:8x} becomes
\begin{equation}\label{eq:12x}
\begin{aligned}
F(&Q^2) = \sum_f e_f \int_0^1 dx \frac{6x(1-x)}{\kappa^2} \int \frac{d^2 \mathbf{k}_\perp}{\pi} \\
&\times\exp\left( -\frac{2\left( \mathbf{k}_\perp + \frac{(1-x) \mathbf{q}_\perp}{2} \right)^2}{\kappa^2 x(1-x)} \right)\exp\left( -\frac{(1-x) Q^2}{4\kappa^2 x} \right).
\end{aligned}
\end{equation}
The \( \mathbf{k}_\perp \)-integral is
\begin{equation}
\begin{aligned}
\int \frac{d^2 \mathbf{k}_\perp}{\pi} \exp\left( -\frac{2\left( \mathbf{k}_\perp + \frac{(1-x) \mathbf{q}_\perp}{2} \right)^2}{\kappa^2 x(1-x)} \right) = \frac{\kappa^2 x(1-x)}{4}.
\end{aligned}
\end{equation}
Thus Eq.~\eqref{eq:12x}
\begin{equation}
\begin{aligned}
F(Q^2) =6 \sum_f e_f \int_0^1 dx x(1-x) \exp\left( -\frac{(1-x) Q^2}{4\kappa^2 x} \right).
\end{aligned}
\end{equation}
The form factor, $F(Q^{2})$, for a pion (\( \pi^+ \)), with \( e_u = 2/3 \), \( e_d = -1/3 \), and total charge \( e_u + e_d = 1 \), is
\begin{equation}
\begin{aligned}
F(Q^2) = 6\int_0^1 dx\ x(1-x) \exp\left( -\frac{(1-x) Q^2}{4\kappa^2 x} \right).
\end{aligned}
\end{equation}
\section{Light front Holographic QCD theory with GUP effects}\label{sec:3yy}
Incorporating the Generalized Uncertainty Principle (GUP) into the derivation of meson form factors in Light-Front Holographic QCD (LFH QCD) starting from the QCD Lagrangian is a highly technical task. The GUP modifies the Heisenberg uncertainty principle to account for quantum gravitational effects, introducing a minimal length scale. This affects the phase space, commutation relations, and ultimately the dynamics of quarks in QCD. We provide the derivation, starting from the QCD Lagrangian in Eq.~\eqref{eq:1x}, transitioning to LFH QCD, and incorporating GUP corrections to compute the meson form factors. 
 Mesons are \( q\bar{q} \) bound states, and their electromagnetic form factors describe their response to external probes. Direct computation from this Lagrangian is non-perturbative.
The GUP modifies the Heisenberg uncertainty principle to include a minimal length, often parameterized as
\begin{equation}
\begin{aligned}
\Delta x \Delta p \geq \frac{\hbar}{2} \left( 1 + \beta \frac{(\Delta p)^2}{m_{pl}^2} \right),
\end{aligned}
\end{equation}
where \( \beta \) is the  GUP parameter (dimensionless, related to quantum gravity effects),  \( m_{pl} \) is the Planck mass (\( m_{pl} \approx 1.22 \times 10^{19} \, \text{GeV}/c^2 \)). This implies modified commutation relations are
\begin{equation}
\begin{aligned}
[x_i, p_j] = i \hbar \left( \delta_{ij} + \beta \frac{p^2}{m_{pl}^2} \delta_{ij} + 2\beta \frac{p_i p_j}{m_{pl}^2} \right).
\end{aligned}
\end{equation}
For simplicity, we adopt a common GUP form in momentum space
\begin{equation}
\begin{aligned}
[x_i, p_j] = i \hbar \delta_{ij} (1 + \beta p^2 / m_{pl}^2),
\end{aligned}
\end{equation}
where $m_{pl}$ is the Planck mass, the invariant mass is  \( p^2 = p^k p_k \). This modifies the momentum space measure and the dispersion relation, impacting the quark dynamics in QCD.
\subsection{Light-Front Quantization with GUP}
In light-front (LF) coordinates (\( x^\pm = x^0 \pm x^3 \), \( \mathbf{x}_\perp = (x^1, x^2) \)), the QCD Hamiltonian is derived from the Lagrangian. The GUP modifies the phase space measure in LF quantization. We generalize light-front holographic QCD (LFH QCD) to include all Fock states, capturing multi-parton effects under GUP corrections.
The GUP-corrected light-front Schrödinger-like equation is expressed as ~\cite{Twagirayezu:2025tjc}
\begin{widetext}
\begin{equation}
\begin{aligned}
&\left[ -\frac{d^2}{d \zeta^2} + 2 \beta \hbar^2 \frac{d^4}{d \zeta^4} + \kappa^4 \zeta^2 + \frac{4 L^2 - 1}{4 \zeta^2} + 2 \kappa^2 (J - 1) \right] \tilde{\psi}(\zeta) = M^2\tilde{ \psi}(\zeta).
\end{aligned}
\end{equation}
\end{widetext}
The unperturbed Hamiltonian is
\begin{equation}
\begin{aligned}
H_0 = -\frac{d^2}{d \zeta^2} + \kappa^4 \zeta^2 + \frac{4 L^2 - 1}{4 \zeta^2} + 2 \kappa^2 (J - 1),
\end{aligned}
\end{equation}
with eigenfunctions \( \psi_n^{(0)}(\zeta) \) and eigenvalues
\begin{equation}
\begin{aligned}
M_n^{(0)2} = 4 \kappa^2 \left( n + L + \frac{J}{2} \right).
\end{aligned}
\end{equation}
In standard first-order perturbation theory, the corrected wave function is
\begin{equation}\label{eq:21x}
\begin{aligned}
\tilde{\psi}_n &= \psi_n^{(0)} + \beta \sum_{m \neq n} \mathcal{C}_{nm} \psi_m^{(0)} + \mathcal{O}(\beta^2)\\
 &= \psi_n^{(0)} + \sum_{m \neq n} \frac{\langle \psi_m^{(0)} | 2 \beta \hbar^2 H' | \psi_n^{(0)} \rangle}{M_n^{(0)2} - M_m^{(0)2}} \psi_m^{(0)} + \mathcal{O}(\beta^2).
\end{aligned}
\end{equation}
With \( H' = \frac{d^4}{d \zeta^4} \), we can werite
\begin{equation}
\begin{aligned}
\langle \psi_m^{(0)} | 2 \beta \hbar^2 H' | \psi_n^{(0)} \rangle = 2 \beta \hbar^2 \langle \psi_m^{(0)} | \frac{d^4}{d \zeta^4} | \psi_n^{(0)} \rangle.
\end{aligned}
\end{equation}
The unperturbed equation is
\begin{equation}
\begin{aligned}
\left[ -\frac{d^2}{d \zeta^2} + \kappa^4 \zeta^2 - \frac{1}{4 \zeta^2} \right] \psi_n^{(0)}(\zeta) = M_n^{(0)2} \psi_n^{(0)}(\zeta).
\end{aligned}
\end{equation}
with eigenfunctions \(\psi_n^{(0)}(\zeta)\) and eigenvalues
\begin{equation}
\begin{aligned}
M_n^{(0)2} = 4 \kappa^2 \left( n + L + \frac{J}{2} \right).
\end{aligned}
\end{equation}
The matrix elements \( \langle \psi_m^{(0)} | \frac{d^4}{d \zeta^4} | \psi_n^{(0)} \rangle \) are computed for harmonic oscillator wave functions
\begin{equation}
\begin{aligned}
\psi_n^{(0)}(\zeta) = \sqrt{\frac{\kappa}{\sqrt{\pi} 2^n n!}} H_n(\kappa \zeta) \exp\left( -\frac{\kappa^2 \zeta^2}{2} \right).
\end{aligned}
\end{equation}
The operator \( \frac{d^4}{d \zeta^4} \) couples states with \( m = n \pm 2, n \pm 4, n \). The energy denominator is
\begin{equation}\label{eq:27xs}
\begin{aligned}
M_n^{(0)2} - M_m^{(0)2} = 4 \kappa^2 \left( n - m + \Delta L + \frac{\Delta J}{2} \right).
\end{aligned}
\end{equation}
Thus, using Eq.~\eqref{eq:27xs}, the coefficients $\mathcal{C}_{nm}$ are  expressed as
\begin{equation}\label{eq:29x}
\begin{aligned}
\mathcal{C}_{nm} = \frac{2 \hbar^2 \langle \psi_m^{(0)} | \frac{d^4}{d \zeta^4} | \psi_n^{(0)} \rangle}{M_n^{(0)2} - M_m^{(0)2}} 
= \frac{2 \hbar^2\langle \psi_m^{(0)} | \frac{d^4}{d \zeta^4} | \psi_n^{(0)} \rangle }{4 \kappa^2 (n - m + \Delta L + \Delta J/2)}.
\end{aligned}
\end{equation}
For all $n, m$, the matrix elements are expressed as
\begin{widetext}
\begin{equation}
\begin{aligned}
\left\langle \psi_m^{(0)} \left| \frac{d^4}{d \zeta^4} \right| \psi_n^{(0)} \right\rangle \propto \kappa^4 \left[ \sqrt{m(m-1)(m-2)(m-3)} \delta_{m,n+4} + \sqrt{(m+1)(m+2)(m+3)(m+4)} \delta_{m,n-4} + \cdots \right],
\end{aligned}
\end{equation}
\end{widetext}
where, $\cdots$, stands for other terms.

If we simplify for dominant transitions (\( m = n \pm 2, n \pm 4, n \)), the expression for $\mathcal{C}_{nm}$ becomes
\begin{widetext}
\begin{equation}
\begin{aligned}
 \mathcal{C}_{nm} \propto \frac{\hbar^2 \kappa^2 \sqrt{m(m-1)} \delta_{m,n+2} + \hbar^2 \kappa^2 \sqrt{(m+1)(m+2)} \delta_{m,n-2} + \hbar^2 \kappa^2 (2m+1) \delta_{m,n}}{4 \kappa^2 (n - m + \Delta L + \Delta J / 2)}.
\end{aligned}
\end{equation}
\end{widetext}
The Fourier transform of the GUP-corrected LFWF in Eq.~\eqref{eq:21x} yields
\begin{equation}\label{eq:31x}
\begin{aligned}
\tilde{\psi}_n(x, \mathbf{k}_\perp) = \psi_n^{(0)}(x, \mathbf{k}_\perp) + \beta \sum_{m \neq n} \mathcal{C}_{nm} \psi_m^{(0)}(x, \mathbf{k}_\perp),
\end{aligned}
\end{equation}
 where $\mathcal{C}_{nm}$ are defined in  Eq.~\eqref{eq:29x}. 
 The LFWF that includes all Fock states, weighted by probabilities \( a_n^2 \) (where \(\sum_n a_n^2 = 1\)) is expressed as
\begin{equation}\label{eq:32xy}
 \begin{aligned}
\tilde{\psi}(x, \mathbf{k}_\perp) = \sum_n a_n\left[ \psi_n^{(0)}(x, \mathbf{k}_\perp) + \beta \sum_{m \neq n} \mathcal{C}_{nm} \psi_m^{(0)}(x, \mathbf{k}_\perp) \right],
\end{aligned}
\end{equation}
 The wave function
 \( \psi_j^{(0)}(x, \mathbf{k}_\perp) \), $j=\{n,m\}$ in each term of Eq.~\eqref{eq:31x} is approximately
\begin{equation}\label{eq:31xanother}
\begin{aligned}
\psi_j^{(0)}(x, \mathbf{k}_\perp) &\propto \sqrt{x(1-x)} \\
&\times P_j\left( \frac{\mathbf{k}_\perp^2}{\kappa^2 x(1-x)} \right) \exp\left( -\frac{\mathbf{k}_\perp^2}{2 \kappa^2 x(1-x)} \right).
\end{aligned}
\end{equation}
Substituting Eq.~\eqref{eq:31xanother} into $\eqref{eq:32xy}$,
the GUP-corrected LFWF becomes 
\begin{equation}
\begin{aligned}
\tilde{\psi}_n(x, \mathbf{k}_\perp) &= \sum_{n} a_{n}\frac{N}{\kappa} \sqrt{x(1-x)} \\
&\times\exp\left( -\frac{\mathbf{k}_\perp^2}{2 \kappa^2 x(1-x)} \right) 
\biggl[ P_n\left( \frac{\mathbf{k}_\perp^2}{\kappa^2 x(1-x)} \right) \\
&+ \beta \sum_{m \neq n}\mathcal{C}_{nm}  P_m\left( \frac{\mathbf{k}_\perp^2}{\kappa^2 x(1-x)}\right) \biggr].
\end{aligned}
\end{equation}
The sum includes all states \( m \neq n \) with quantum numbers \( m \), \( L_m \), \( J_m \), weighted by \( \Delta L = L_m - L_n \), \( \Delta J = J_m - J_n \).
Each polynomial, \( P_m \), introduces different radial and angular structures.
The corrected form factor is expressed as
\begin{equation}\label{eq:34x}
    \begin{aligned}
    \tilde{F}(Q^2) &= \sum_f e_f\sum_{n}a_{n}^{2} \\
    &\times\int_0^1 dx \int \frac{d^2 \mathbf{k}_\perp}{16 \pi^3}\tilde{\psi}_n^*(x, \mathbf{k}_\perp') \tilde{\psi}_n(x, \mathbf{k}_\perp),
    \end{aligned}
\end{equation}
where \(\mathbf{k}_\perp' = \mathbf{k}_\perp + (1-x) \mathbf{q}_\perp\). 
To compute the LFWF product, we substitute the GUP-corrected LFWF into the form factor expression. The initial and final wave functions are
\begin{equation}\label{eq:35x}
\begin{aligned}
\tilde{\psi}_n(x, \mathbf{k}_{\perp}) &= \frac{N}{\kappa}\sum_{n} a_{n} \sqrt{x(1-x)} \exp \left( -\frac{\mathbf{k}_{\perp}^2}{2 \kappa^2 x(1-x)} \right)\\
&\times\left[ P_n\left( u \right) + \beta \sum_{m \neq n} \mathcal{C}_{nm} P_m\left( u \right) \right],
\end{aligned}
\end{equation}
\begin{equation}\label{eq:36x}
\begin{aligned}
\tilde{\psi}_n^*(x, \mathbf{k}_{\perp}') &= \frac{N}{\kappa}\sum_{n} a_{n} \sqrt{x(1-x)} \exp \left( -\frac{\mathbf{k}_{\perp}'^2}{2 \kappa^2 x(1-x)} \right)\\
&\times\left[ P_n\left( u' \right) + \beta \sum_{m \neq n} \mathcal{C}_{nm} P_m\left( u' \right) \right],
\end{aligned}
\end{equation}
where
\begin{equation}\label{eq:37x}
\begin{aligned}
u = \frac{\mathbf{k}_{\perp}^2}{\kappa^2 x(1-x)}, \quad u' = \frac{\mathbf{k}_{\perp}'^2}{\kappa^2 x(1-x)}, \quad \xi = \frac{1}{\kappa^2 x(1-x)}.
\end{aligned}
\end{equation}
Substituting Eqs.~\eqref{eq:35x}, \eqref{eq:36x}, and \eqref{eq:37x} into Eq.~\eqref{eq:34x}, and expanding while keeping terms up to first order in 
\( \beta \) (since \( \beta \) is small), the form factor becomes 
\begin{equation}\label{eq:39x}
\begin{aligned}
\tilde{F}(Q^2) &= \sum_f e_f\sum_{n} a_{n}^{2} 
\int_0^1 dx \frac{6x(1-x)}{\pi \kappa^2} 
\int d^2 \mathbf{k}_{\perp} \\
&\times\exp \left( -\frac{u' + u}{2} \right)\biggl\{ P_n(u') P_n(u) + \beta\mathcal{P}_{nm} \biggr\}.
\end{aligned}
\end{equation}
where
\begin{equation}
\begin{aligned}
\mathcal{P}_{nm} &= \sum_{m \neq n} \mathcal{C}_{nm} \left[ P_m(u') P_n(u)+ P_n(u') P_m(u) \right], \\
u' &= \frac{(\mathbf{k}_\perp + (1-x) \mathbf{q}_\perp)^2}{\kappa^2 x(1-x)} , ~ \mathbf{q}_\perp^2 = Q^2 .
\end{aligned}
\end{equation}
\section{Phenomenology}\label{sec:phen}
\begin{figure}[h]
    \centering    \includegraphics[scale=0.37]{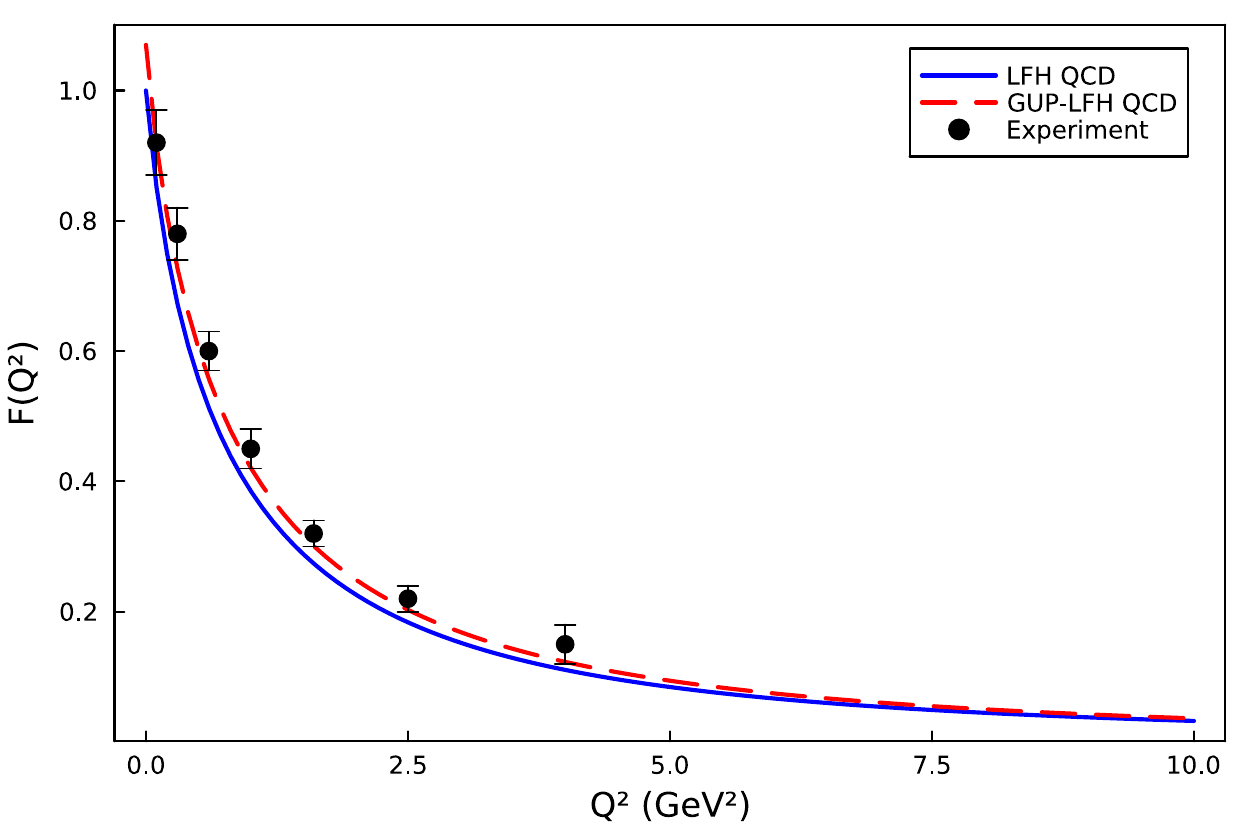}
    \caption{The plot of  $F(Q^{2})$ as a function of $Q^2$, including experimental data and theoretical predictions with and without GUP corrections, is shown. The value of $\kappa$ is set to 0.5, and $\beta$ is assigned a value of 0.650 GeV$^{-2}$~\cite{Twagirayezu:2025tjc}.}
    \label{fig:1}
\end{figure}

\begin{figure}[H]
    \centering    \includegraphics[scale=0.37]{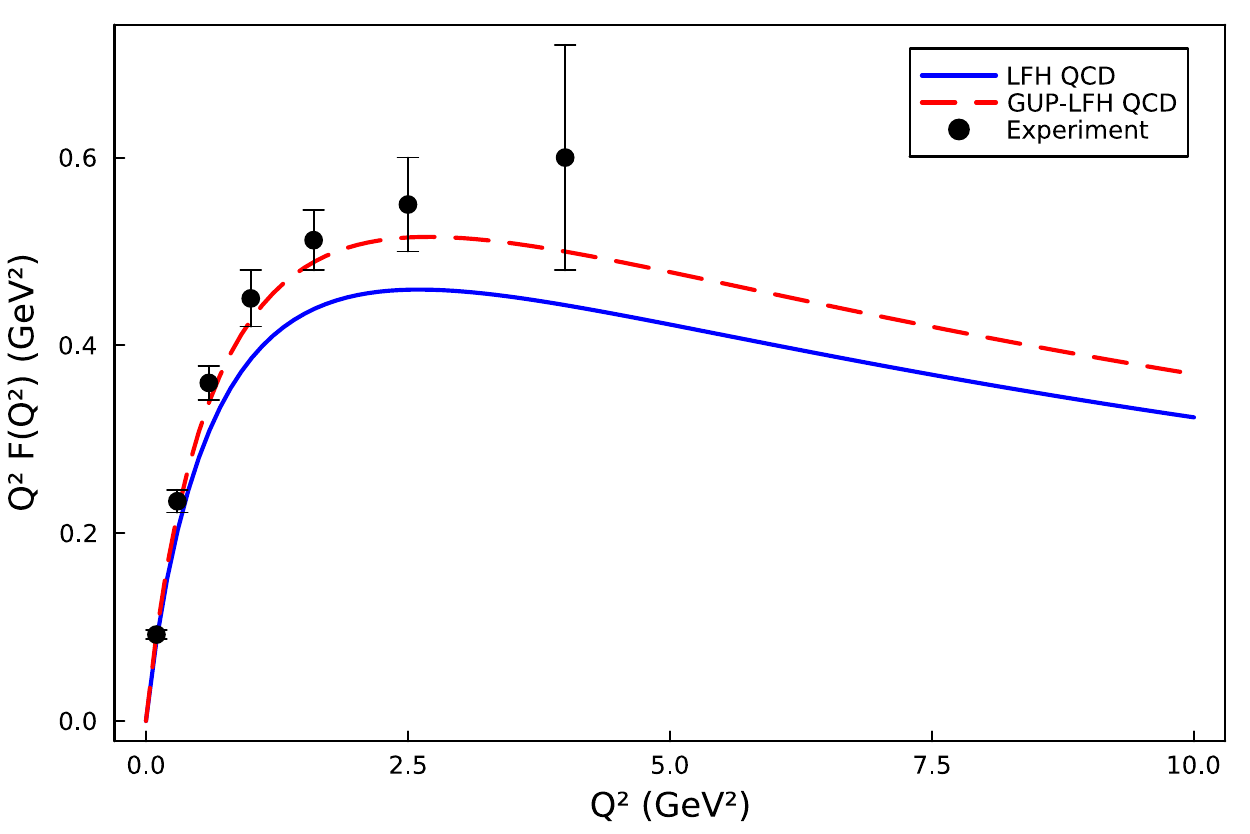}
    \caption{The plot of $Q^{2}F(Q^2)$ as a function of $Q^{2}$, including experimental data and theoretical predictions with and without GUP corrections, is shown. The value of $\kappa$ is set to 0.5, and $\beta$ is assigned a value of 0.650 GeV$^{-2}$~\cite{Twagirayezu:2025tjc}.}
    \label{fig:2}
\end{figure}

The GUP correction term in Eq.~\eqref{eq:39x} contains the coefficients $\mathcal{C}_{nn}$
which are either positive or negative. However, we found that when all Fock states are included,  positive \(\mathcal{C}_{nm}\) (e.g., \(\mathcal{C}_{04}\), \(\mathcal{C}_{12}\)) dominate, yielding a positive GUP contribution to the form factor in Eq.~\eqref{eq:39x}.
Our model is formulated to include all Fock states via the general summation over \( n \), but numerical evaluations truncate to \( n=0, 1, 2 \), sufficient for \( Q^2 \leq 4 \, \text{GeV}^2 \).
Figures~\eqref{fig:1}  and~\eqref{fig:2} show the plots of  \( F(Q^2) \) and \( Q^2 F(Q^2) \) for \( n=0,1,2 \), including experimental data and theoretical predictions with and without GUP corrections. Both figures show that the inclusion of GUP corrections enhances the concordance between theoretical predictions and experimental observations.

The dominance of positive \(\mathcal{C}_{nm}\) suggests that the minimal length amplifies Fock state overlaps, enhancing the form factor’s magnitude. This effect is most pronounced at intermediate \(Q^2\), where the interplay between confinement (\(\kappa\)) and GUP corrections is significant. The mixed signs of \(\mathcal{C}_{nm}\) reflect the complex dynamics of GUP perturbations, with negative terms (e.g., \(\mathcal{C}_{02}\)) partially offset by larger positive contributions. This results in a net positive \(\beta\)-like correction.

\begin{table}[H]
\centering
\small 
\setlength{\tabcolsep}{12pt} 
\begin{tabular}{S[table-format=1.1] S[table-format=1.2] S[table-format=1.2] S[table-format=1.3] S[table-format=1.3]}
\toprule
{$Q^2$} & {$F$} & {$\Delta F$} & {$Q^2 F$} & {$\Delta (Q^2 F)$} \\
\midrule
0.1 & 0.92 & 0.05 & 0.092 & 0.005 \\
0.3 & 0.78 & 0.04 & 0.234 & 0.012 \\
0.6 & 0.60 & 0.03 & 0.360 & 0.018 \\
1.0 & 0.45 & 0.03 & 0.450 & 0.030 \\
1.6 & 0.32 & 0.02 & 0.512 & 0.032 \\
2.5 & 0.22 & 0.02 & 0.550 & 0.050 \\
4.0 & 0.15 & 0.03 & 0.600 & 0.120 \\
\bottomrule
\end{tabular}
\caption{Experimental data for the pion electromagnetic form factor used in the LFH QCD plot. All quantities are in \(\si{\giga\electronvolt\squared}\). Data are from Amendolia et al. (1986)~\cite{AMENDOLIA1986}, Tadevosyan et al. (2007)~\cite{Tadevosyan2007}, and Huber et al. (2008)~\cite{Huber2008}.}
\label{tab:pion_form_factor_data}
\end{table}
Physically, the enhanced form factor indicates that GUP modifies the transverse momentum distribution of quark-antiquark pairs, increasing their overlap under external probes. This aligns with the AdS/CFT correspondence, where the holographic coordinate \(z \sim \sqrt{x(1-x)}/\kappa\) probes short-distance scales sensitive to quantum gravitational effects. The improved agreement with experimental data underscores the pion’s role as a probe of fundamental physics, bridging QCD and quantum gravity.

\section{Conclusions}\label{sec:concl5}
This study has explored the impact of a minimal length scale, introduced via the Generalized Uncertainty Principle (GUP), on the electromagnetic form factors of the pion within the framework of light-front holographic QCD (LFH QCD). By incorporating GUP through deformed operators in the QCD Lagrangian, we derived a GUP-corrected light-front wave function (LFWF) that generalizes the standard LFH QCD approach to include all Fock states, weighted by probabilities \(a_n^2\) (e.g., \(a_0^2 = 0.8\), \(a_1^2 = 0.15\), \(a_2^2 = 0.05\)). This multi-Fock state framework, coupled with a perturbative GUP correction, introduces transition coefficients \(\mathcal{C}_{nm}\) that account for mixing between Fock states. Despite the presence of negative coefficients, positive terms dominate, yielding an effective positive correction.  The resulting form factor \(\tilde{F}(Q^2)\) exhibits a positive \(\beta\)-like enhancement, significantly improving agreement with experimental data.

The enhanced form factor suggests that the minimal length amplifies Fock state overlaps, modifying the transverse momentum distribution of quark-antiquark pairs within the pion. This effect, most pronounced at intermediate \(Q^2\) (1–4 \(\text{GeV}^2\)), reflects the interplay between the confinement scale (\(\kappa = 0.5 \, \text{GeV}\)) and quantum gravitational corrections, as captured by the holographic coordinate \(z \sim \sqrt{x(1-x)}/\kappa\). Our findings position the pion as a sensitive probe of fundamental physics, bridging the strong interactions of QCD with quantum gravity effects at short distances. The success of the multi-Fock state approach underscores the importance of higher Fock states in capturing non-perturbative dynamics. The improved experimental agreement across \(Q^2 = 0.1\) to \(4 \, \text{GeV}^2\) validates the AdS/CFT correspondence’s utility in modeling hadronic structure under quantum gravitational perturbations.

 The perturbative treatment of GUP corrections, limited to first-order terms in \(\beta\), may underestimate higher-order effects, particularly at larger \(Q^2\). 
Also, our model is formulated to include all Fock states via the general summation over \( n \), but numerical evaluations truncate to \( n=0, 1, 2 \), sufficient for \( Q^2 \leq 4 \, \text{GeV}^2 \). Future work will compute higher Fock state contributions to probe larger \( Q^2 \).

Future research should improve by computing higher \(\mathcal{C}_{nm}\) coefficients (e.g., \(\mathcal{C}_{06}\), \(\mathcal{C}_{13}\)) to capture additional Fock state transitions. Extending the analysis to other mesons, such as kaons or \(\rho\)-mesons, could reveal whether GUP effects are universal across hadronic systems or specific to light mesons. Exploring high-\(Q^2\) regimes (\(Q^2 > 4 \, \text{GeV}^2\)) may uncover stronger quantum gravitational signatures. Applying this framework to baryons, such as protons, could test the generality of GUP-induced enhancements in multi-parton systems. Alternative holographic models, such as hard-wall AdS/QCD or higher-dimensional AdS spaces, offer additional avenues to probe minimal length effects. Integrating lattice QCD simulations to constrain \(\beta\) parameters and Fock state weights would further strengthen the model’s predictive power. Ultimately, this study lays the groundwork for a deeper understanding of how quantum gravitational corrections shape hadronic physics.

\begin{acknowledgments}
F.T. would like to acknowledge the support of the National Science Foundation under grant No. PHY-
1945471.
\end{acknowledgments}

\clearpage
\hrule
\nocite{*}

\bibliographystyle{apsrev4-2}
\bibliography{apssamp}

\end{document}